\documentclass[twocolumn]{aastex631} 
\NewPageAfterKeywords
\pdfoutput=1
\newcommand{\umich}{University of Michigan, Department of Astronomy, 1085 South University, Ann Arbor MI, 48109, USA}

\newcommand\tbd[1]{{\color{red}#1}}

\newcommand\editsone{}
\newcommand\editstwo{}

\newcommand\Msun{$M_{\odot}$} %
\newcommand\Rsun{$R_{\odot}$} %
\newcommand\Lsun{$L_{\odot}$} %

\usepackage{amsmath}
\usepackage[flushmargin]{footmisc}
\newcommand{\panelcaption}[1]{%
  \vspace{1mm}
  {\centering\small #1\par}
  \vspace{2mm}
}
\usepackage{hyperref}
\begin{document}

\title{Assessing Planetary Stability and Long-Term Habitability in Nearby Stellar Binaries: 70 Oph, 36 Oph, $\gamma$ Leo}

\author[0000-0003-4287-004X]{Alyssa R. Jankowski}
\affiliation{University of Wisconsin - Madison Department of Astronomy, 475 N. Charter St. Madison, WI 53706, USA}
\affiliation{Wisconsin Center for Origins Research, 6515 Sterling Hall, 475 N Charter St, Madison, WI 53706, USA}

\author[0000-0002-7733-4522]{Juliette Becker}
\affiliation{University of Wisconsin - Madison Department of Astronomy, 475 N. Charter St. Madison, WI 53706, USA}
\affiliation{Wisconsin Center for Origins Research, 6515 Sterling Hall, 475 N Charter St, Madison, WI 53706, USA}

\author[0000-0002-2361-5812]{Catherine A. Clark}
\affil{NASA Exoplanet Science Institute, Caltech/IPAC, 1200 E. California Blvd., MC 100-22, Pasadena, CA 91125, USA}

\author[0000-0003-2008-1488]{Eric E. Mamajek}
\affiliation{Jet Propulsion Laboratory, California Institute of Technology, 4800 Oak Grove Dr., Pasadena, CA 91109, USA}

\author[0000-0002-8035-4778]{Jessie L. Christiansen}
\affiliation{NASA Exoplanet Science Institute, IPAC, MS 100-22, Caltech, 1200 E. California Blvd, Pasadena, CA 91125}

\author[0000-0002-6845-9702]{Yiting Li}
\affiliation{\umich}

\author[0000-0002-9408-8925]{Eduardo Bendek}
\affiliation{NASA Ames Research Center, Moffett Field, CA, 94035}

\author[0000-0001-8170-7072]{Daniella Gagliuffi}
\affiliation{Department of Physics \& Astronomy, Amherst College, 25 East Drive, Amherst, MA 01003, USA}

\author[0000-0002-0078-5288]{Mark R. Giovinazzi}
\affiliation{Department of Physics \& Astronomy, Amherst College, 25 East Drive, Amherst, MA 01003, USA}

\author[0009-0000-4050-4172]{Simone Lilavois}
\affiliation{Department of Physics \& Astronomy, Amherst College, 25 East Drive, Amherst, MA 01003, USA}

\author[0009-0007-5640-0061]{Ryan Purviance}
\affiliation{Department of Physics \& Astronomy, Amherst College, 25 East Drive, Amherst, MA 01003, USA}

\begin{abstract}

\editsone{Binary stars are common and have the potential to host habitable planets, which may reside in more complex habitable zones as compared to planets orbiting single stars.} In this work, we use numerical simulations to assess the possibility that bright, nearby stellar multiples 36 Oph, 70 Oph, and $\gamma$ Leo could host habitable planets. \editsone{We find that for the 36 Oph A/B system and for the 70 Oph A/B system, the stars can support planets residing in permanently habitable zones with low ejection rates and moderate eccentricity oscillations.} The \editstwo{habitable zones around the} red giants in the $\gamma$ Leo system exhibit severe dynamical instability due to the high binary eccentricity, eliminating the habitable zones around both stars.  \editsone{In these two systems, we find that planets in the habitable zone with orbits coplanar to that of the binary become uninhabitable due to interactions with the binary only $1.5\% - 1.8\%$ of the time, while planets with orbits 45 degrees misaligned to the plane of the binary experience larger oscillations in orbital eccentricity and as a result become uninhabitable $4.8\% - 5.4\%$ of the time.} Our results identify 36 Oph and 70 Oph as promising targets for future missions such as the Habitable Worlds Observatory and SHERA, while suggesting that the stars in $\gamma$ Leo are unlikely to host any habitable planets. Our methods can be applied more generally to other binary stellar systems to refine target lists for upcoming habitable planet searches. 
\end{abstract}

\tbd{}

\section{Introduction} \label{sec:intro}
Stellar multiples are common; roughly half of all Sun-like stars have at least one stellar companion \editsone{\citep[e.g.,][and references therein]{Raghavan2010, Moe2017, Offner2023}.} 
\editsone{Despite their commonality, there are still challenges to characterizing and analyzing these systems. 
Due to the difficulty of detecting planets in stellar binaries and choices in planet-hunting survey design \citep[Kepler, for example, had a more difficult time finding planets around binaries due to both detection bias and pipeline design choices;][]{Borucki2010}, a comparatively low number \citep{Thebault2025} of known exoplanets reside in stellar multiples\footnote{As of 5/13/2026, 5724 out of 6286 planets in the IPAC NASA Exoplanet Archive Database's Confirmed Planets Table \citep{Christiansen2025} are in single-star systems.}.} 
\editsone{Planet detection efforts in stellar multiples are limited by dilution of signals in both transit \citep{Ciardi2015} and radial velocity data \citep{Rucinski2002}, contamination of spectra \citep{Furlan2020}, and added signal complexity compared to a single host star.}
\editsone{Despite these challenges, numerous studies over the past decade have examined planetary habitability in stellar binaries \citep[e.g.,][]{Haghighipour2006, Eggl2012, Kaltenegger2013, Forgan2012, Forgan2015, Quarles2022}, exploring how variable irradiation and dynamical evolution can still permit habitable conditions.}

\editsone{Observational} studies suggest that planet formation is significantly less efficient in close binary systems with $a<10-50$ AU \citep{WangFischer2014, WangXie2014, Wang2015, Kraus2016, Moe2021, Fontanive2021}. Gravitational perturbations from a stellar companion can alter the surface density or velocity profile of the protoplanetary disk \citep{Batygin2011,Harris2012, Akeson2019, Silsbee2021}, and the long-term dynamical evolution of planetary orbits in such environments can be complex \citep{Bazso2017, Quarles2019, Quarles2020}, \editsone{as there are increased risks of orbital instability or ejection compared to single-star systems \citep{Holman1999, Marzari2016}.}

For \editsone{binaries} with larger periastron distances \editsone{($a>50$ AU)}, planet formation in the habitable zone (HZ) is more likely \citep{Haghighipour2007}. 
However, even in such cases, the presence of a companion can significantly impact a planet’s long-term habitability through dynamical effects \citep{Eggl2012, Georgakarakos2018}.
Secular perturbations \citep{MurrayDermott1999} and Kozai-Lidov cycles \citep{Kozai1962, Lidov1962} can lead to variations in orbital eccentricity and inclination, which in turn affect a planet’s climate stability and surface conditions over geological timescales \citep[e.g.,][]{Vervoort2024}. 
Understanding habitability in binary or higher-order stellar systems therefore requires a more nuanced approach that considers not only the planet's location within the habitable zone, but also the evolving gravitational interactions between the stars in the system.

\editsone{Next-generation missions like the Habitable Worlds Observatory \citep[HWO;][]{Gaudi2020,Feinberg2024} and Searching for Habitable Exoplanets with Relative Astrometry \citep[SHERA, a Small Explorer-class astrometry mission which would study the nearest binary systems to search for habitable planets;][]{ChristiansenINPREP} will target a small select group of nearby systems for in-depth characterization \citep{Mamajek2024,Tuchow2025}.} 
{With this in mind, it is increasingly important to identify which multi-star systems might host habitable planets, and which have stellar orbits that exclude stable habitable planets.} 

In this work, we present detailed case studies of the habitable zones of three nearby \editsone{($d$ $\lesssim$ 40\,pc)} stellar multiples: 36 Ophiuchi, 70 Ophiuchi, and $\gamma$ Leonis. 
\editsone{All three systems are relatively bright,} making them possible observational targets for future missions such as the Roman Space Telescope Coronagraph Instrument \citep[CGI;][]{bailey2023}, HWO \citep{Mamajek2024}, and SHERA \citep{ChristiansenINPREP}. 
\editstwo{In particular, these systems are well suited for the proposed SHERA Small Explorer-class mission, which is designed to search for habitable planets through relative astrometry in nearby binary systems, \emph{provided they are capable of hosting habitable planets}.}
Our goal is to determine whether these systems could host yet-undiscovered habitable planets given their currently understood orbits, and subsequently determine if these systems should be prioritized for future HWO/SHERA observations.

In Section \ref{sec:data}, we describe the data and literature solutions that we use to assess the systems' currently accepted orbital parameters.
In Section \ref{sec:sims}, we describe the numerical simulations used to assess the habitability prospects of planets in these systems.
In Section \ref{sec:discussion}, we explore the implications of our findings and offer recommendations for future observational strategies.
Finally, we conclude in Section \ref{sec:conclusion} with a summary of our results. 

\section{Data Sources and System Parameters} \label{sec:data}
For all of the systems considered in this work, there are decades of astrometric and radial velocity observations that have allowed previous authors to refine the binary orbits. In this section, we describe the solutions \editsone{we adopt in this work} for each binary and their sources, \editsone{along with the calculation for the nominal habitable zone location(s) in each system.}

\subsection{Habitable Zone Calculation}
\label{sec:habit}
\editsone{For binary stars separated by several hundred AU, the secondary star's insolation has a negligible impact on a planet's climate \citep{Kaltenegger2013}. As such, the calculation for our target stars' habitable zones can be assumed to be that of a single star. To determine the habitable zone for each of our stars, we compute the flux using Equation 1 from \citet{Bolmont2016} \citep[which includes the effect of orbital eccentricity in setting the flux; see also][]{Adams2006ecc, Gallo2024}:
\begin{equation}
    F_p = \frac{L_{\star}}{4\pi a_p^2\sqrt{1-e_p^2}}
    \label{eq:Bolmont2016}
\end{equation}
where $L_{\star}$ is stellar luminosity, $a_p$ is semi-major axis of the planet's orbit, and $e_p$ is the planetary eccentricity. To choose the fiducial location of the habitable zone, we assume that $e_p=0$, and set our inner and outer limits of the habitable zone such that they receive $F_p = 1.7 F_{\Earth}$ and $F_p = 0.3 F_{\Earth}$, respectively (assuming that $F_{\Earth}=1361\ W/m^2$, following equation 2 of \citealt{Bolmont2016}). We will use this equation (including the eccentricity correction for variable $e_p$) to test whether our simulated planets remain in their habitable zone for the entirety of the simulations, which we discuss further in Section \ref{sec:sims}.}

\begin{deluxetable}{lcc}[htb!]
\tabletypesize{\scriptsize}
\tablecaption{Properties of the 36 Oph System \label{tab:36OphParams}}
\tablehead{\colhead{Parameter} & \colhead{Value} & \colhead{Source} }
\startdata
\hline
\multicolumn{3}{c}{36 Oph A Properties}\\
\hline
$\mathrm{M}_*$ (\Msun) & $0.803_{-0.031}^{+0.032}$ & \citet{NEWORBITS} \\
R$_*$ (\Rsun) & $0.728 \pm 0.051$ & \citet{Hardegree2023}\\ 
$L_{*}$ $(L_{\odot})$ & $0.326 \pm 0.00884$ & \citet{Hardegree2023} \\
Inner HZ (AU) & $0.36$ & This Paper\\
Outer HZ (AU) & $0.87$ & This Paper
\\ 
\hline
\multicolumn{3}{c}{36 Oph B Properties}\\
\hline
$\mathrm{M}_*$ (\Msun)  & $0.867\pm 0.033$  & \citet{NEWORBITS}\\
R$_*$ (\Rsun) & $0.718 \pm 0.05$ & \citet{Hardegree2023}\\ 
$L_{*}$ $(L_{\odot})$ & $0.328 \pm 0.00962$ & \citet{Hardegree2023} \\ 
Inner HZ (AU) & $0.43$ & This Paper\\
Outer HZ (AU) & $1.03$ & This Paper\\
\hline
\multicolumn{3}{c}{36 Oph A/B Binary Orbital Properties}\\
\hline
$P_{orb}$ (yr)  & $501_{-10}^{+14}$ & \citet{NEWORBITS}\\
a (AU) & $74.9_{-1.2}^{+1.5}$ & \citet{NEWORBITS}\\
e & $0.8999_{-0.0033}^{+0.0032}$ & \citet{NEWORBITS}\\
\enddata
\end{deluxetable}

\subsection{36 Ophiuchi}
36 Ophiuchi (hereafter referred to as 36 Oph)\footnote{The 36 Oph A/B binary is also often \editsone{identified} in the double star literature as SHJ 243, ADS 10417, WDS J17153-2636AB, and HIP 84405. 36 Oph A is HD 155886 and 36 Oph B is HD 155885.} is a hierarchical triple-star system. At its core lies a closely spaced binary consisting of two K dwarfs \citep[36 Oph A/B;][]{Irwin1996,Luck2017}, with a very distant companion (36 Oph C) at $\sim~4400$ AU. In our subsequent analysis, we consider only the orbit of 36 Oph A/B, \editsone{we assume that 36 Oph C is sufficiently distant and thus dynamically decoupled from the inner binary. We do note that} \citet{Painter2025} reported an astrometric acceleration in 36 Oph C that cannot be explained by the known system components -- suggesting the possibility of an additional undiscovered companion.


\editsone{Previous literature solutions for 36 Oph were reported in \citet{Irwin1996} and \citet{Izmailov2025} (see a discussion of these solutions in Appendix \ref{app:36oph}). }
\editsone{In this work, we use an updated orbit, derived in \citet{NEWORBITS}, in which the authors jointly use the most up-to-date and historical positional measurements, radial velocity measurements, and an observed proper motion anomaly between Hipparcos and Gaia. The relevant parameters of this orbit are reproduced in Table \ref{tab:36OphParams}.}
\editstwo{For the masses of 36 Oph A/B, we use the values derived in \citet{NEWORBITS}, which we reproduce in Table \ref{tab:36OphParams}. \editsone{We adopt the bolometric luminosities and radii for the pair from \citet{Hardegree2023}.}}


Regarding planetary companions in the 36 Oph system, \citet{Wittenmyer2006} previously searched for evidence of planetary companions around 36 Oph A, and found no evidence of companions larger than $0.19\,M_J$ at 0.05 AU (with correspondingly larger lower-mass limits at larger semi-major axes). However, the data precision was insufficient to directly test for the existence of smaller planets.

\begin{figure}
    \centering
    \includegraphics[width=\linewidth]{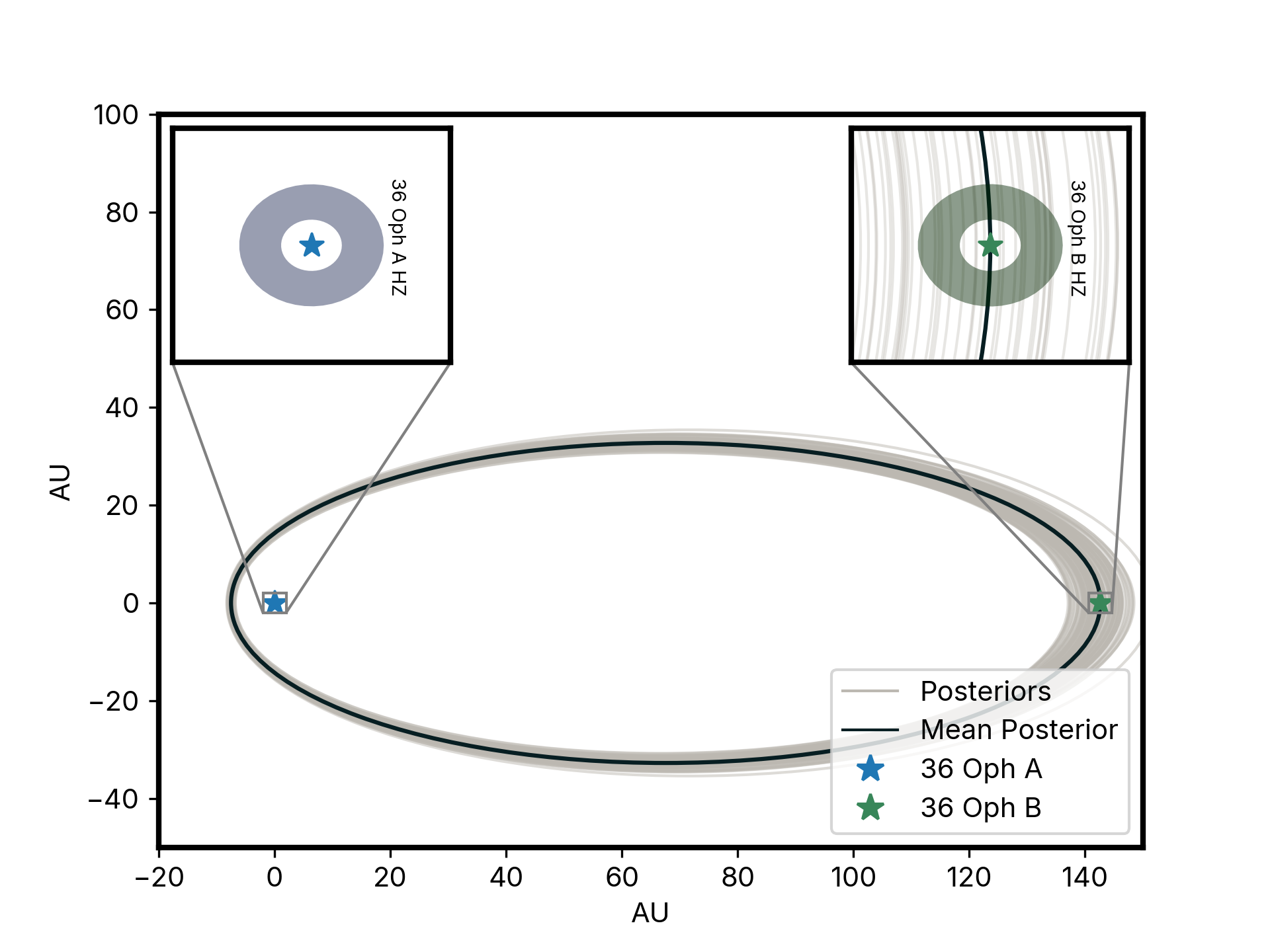}
    \caption{A visual representation of the orbit posteriors for 36 Oph. We fix 36 Oph A at the origin, and include inset plots, which show magnified regions of the individual stellar systems and their habitable zones (shaded blue for 36 Oph A and green for 36 Oph B). \editsone{Note that the orbital posteriors shown are symmetric, and that it would be equally valid to fix 36 Oph B at the origin. In either case, the stellar orbit does not cross the habitable zone of either star; the apparent overlap of the binary orbit with the habitable zone of 36 Oph B is purely a consequence of the visualization. \editstwo{Since we do not model it in our simulations, we do not show 36 Oph C in this plot.}}}
    \label{fig:36OphOrbitPlot}
\end{figure}

\begin{figure}
    \centering
    \includegraphics[width=\linewidth]{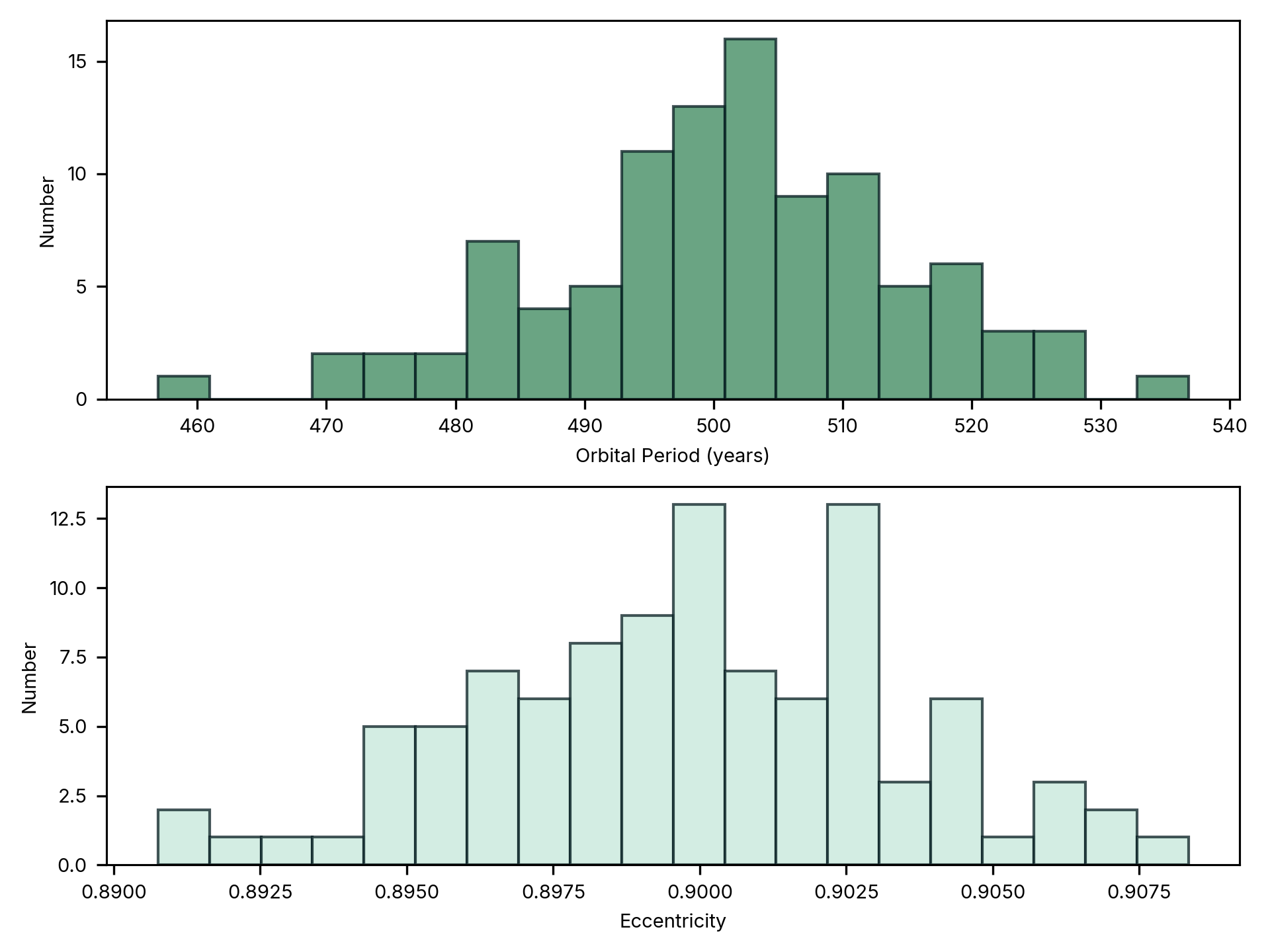}
    \caption{Histograms showing the orbit posteriors for 36 Oph from \citet{NEWORBITS}.}
    \label{fig:36OphOrbitHists}
\end{figure}

\editsone{
Figure \ref{fig:36OphOrbitPlot} presents a visual representation of random samples of the orbit of the 36 Oph system drawn from the range of posteriors from \citet{NEWORBITS}. 
Figure \ref{fig:36OphOrbitHists} shows these samples as a set of histograms showing the distribution of binary orbit parameters. 
We summarize the physical properties, orbital properties, and habitable zones for the 36 Oph system in Table \ref{tab:36OphParams}.}
 \\

\subsection{70 Ophiuchi}
70 Ophiuchi (hereafter referred to as 70 Oph)\footnote{70 Oph is also commonly identified in the double star literature as STF 2272AB, ADS 11046AB, WDS J18055+0230AB, and HIP 88601. 70 Oph A is HD 165341A and 70 Oph B is HD 165341B.} is a binary composed of two K dwarfs: 70 Oph A/B \citep[e.g.,][]{Eggenberger2008}. There are currently no known tertiary stellar companions or planets orbiting either star, \editsone{but this system has a geometry that is broadly of interest. The $\alpha$ Centauri A system, long noted as a good target for searches for habitable planets, has been identified as having a dynamic and variable habitable zone \citep{Forgan2012, Quarles2018}. The system was recently identified to host a planet candidate in the habitable zone \citep{Beichman2025, Sanghi2025}. As a result and by analogy, the 70 Oph binary system could be a promising target for future searches for planets in the habitable zone, as long as the dynamical evolution caused by the binary does not preclude habitable conditions on planets in the habitable zone. } 

\editsone{For this system, we adopt the binary orbit solution reported by \citet{Li2026}. This solution utilizes the most recent visual and radial velocity measurements found in the literature, along with archival data. The authors jointly fit the relative astrometry, absolute astrometry, and multi-instrument radial velocity data using an MCMC sampler.}
We adopt the measurements of the stellar masses from \citet{Piccotti2020}. 
\editsone{For the stellar radii, we adopt \citet{Boyajian2012}. For the effective temperature of 70 Oph A, we use \citet{Soubiran2022}, while for the effective temperature of 70 Oph B, we adopt the value from \citet{Boyajian2012}, as a value for the effective temperature of 70 Oph B was not derived in \citet{Soubiran2022}. For the stellar luminosities, we adopt the values used in \citet{Li2026}, which originate from \citet{Eggenberger2008}.}
\begin{deluxetable}{lcc}[htb!]
\tablecaption{\editsone{Properties of the 70 Oph System}\label{tab:70OphParams}}
\tablehead{\colhead{Parameter} & \colhead{Value} & \colhead{Source} }
\startdata
\hline
\multicolumn{3}{c}{70 Oph A Properties}\\
\hline
$\mathrm{M}_*$ (\Msun)  & $0.8827\pm0.004$ & \citet{Li2026} \\
R$_*$ (\Rsun) & $0.8310\pm 0.0044$ & \citet{Boyajian2012}\\ 
L$_*$ (\Lsun) & $0.53\pm0.02$ & \citet{Li2026}\\
$T_{\rm eff}$ (K) & \editsone{$5298\pm32$} & \editsone{\citet{Soubiran2022}}\\
Inner HZ Radius (AU) & $0.55$ & This Paper\\
Outer HZ Radius (AU) & $1.32$ & This Paper
\\ 
\hline
\multicolumn{3}{c}{70 Oph B Properties}\\
\hline
$\mathrm{M}_*$ (\Msun) & $0.7319\pm0.0031$ & \citet{Li2026} \\
R$_*$ (\Rsun) & $0.6697\pm0.0089$ & \citet{Boyajian2012}\\ 
L$_*$ (\Lsun) & $0.15\pm0.02$ & \citet{Li2026}\\
$T_{\rm eff}$ (K) & $4393\pm149$ & \citet{Boyajian2012}\\
Inner HZ Radius (AU) & $0.29$ & This Paper\\
Outer HZ Radius (AU) & $0.70$ & This Paper\\
\hline
\multicolumn{3}{c}{\editsone{70 Oph Stars' Mean Orbital Properties}}\\
\hline
$P_{orb}$ (yr)  & $88.126\pm 0.010$ & \citet{Li2026} \\
a (AU) & $23.232_{-0.024}^{+0.025}$ & \citet{Li2026}\\
e & $0.50015_{-0.00016}^{+0.00017}$ & \citet{Li2026}
\enddata
\end{deluxetable}

\begin{figure}
    \centering
    \includegraphics[width=\linewidth]{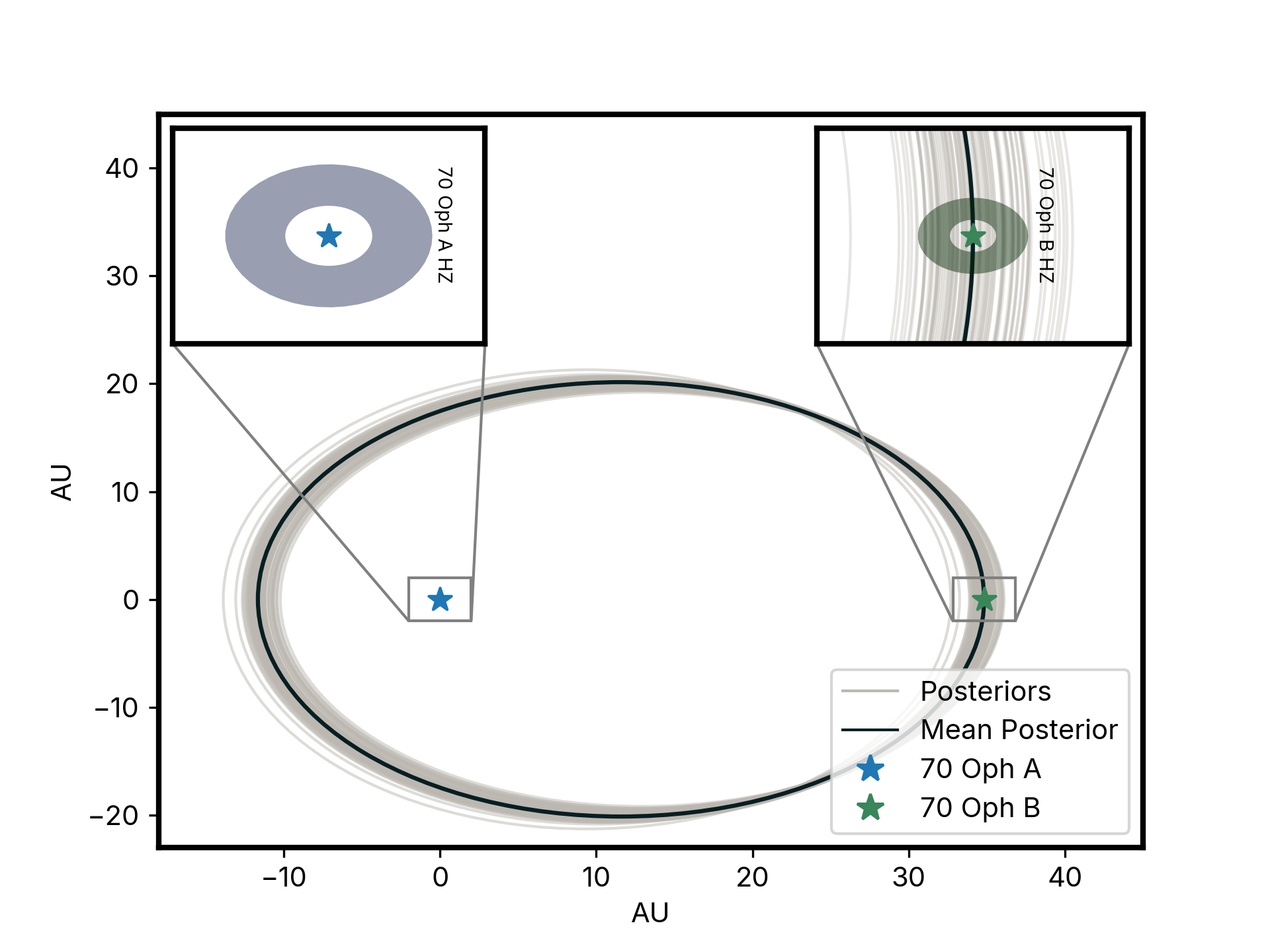}
    \caption{A visual representation of the orbit posteriors for 70 Oph. We fix 70 Oph A at the origin, and include inset plots, which show magnified regions of the individual stellar systems and their habitable zones (shaded blue for 70 Oph A and green for 70 Oph B). \editsone{As for 36 Oph, the apparent overlap with the habitable zone of 70 Oph B is purely a consequence of the visualization.}}
    \label{fig:70OphOrbitPlot}
\end{figure}

\begin{figure}
    \centering
    \includegraphics[width=\linewidth]{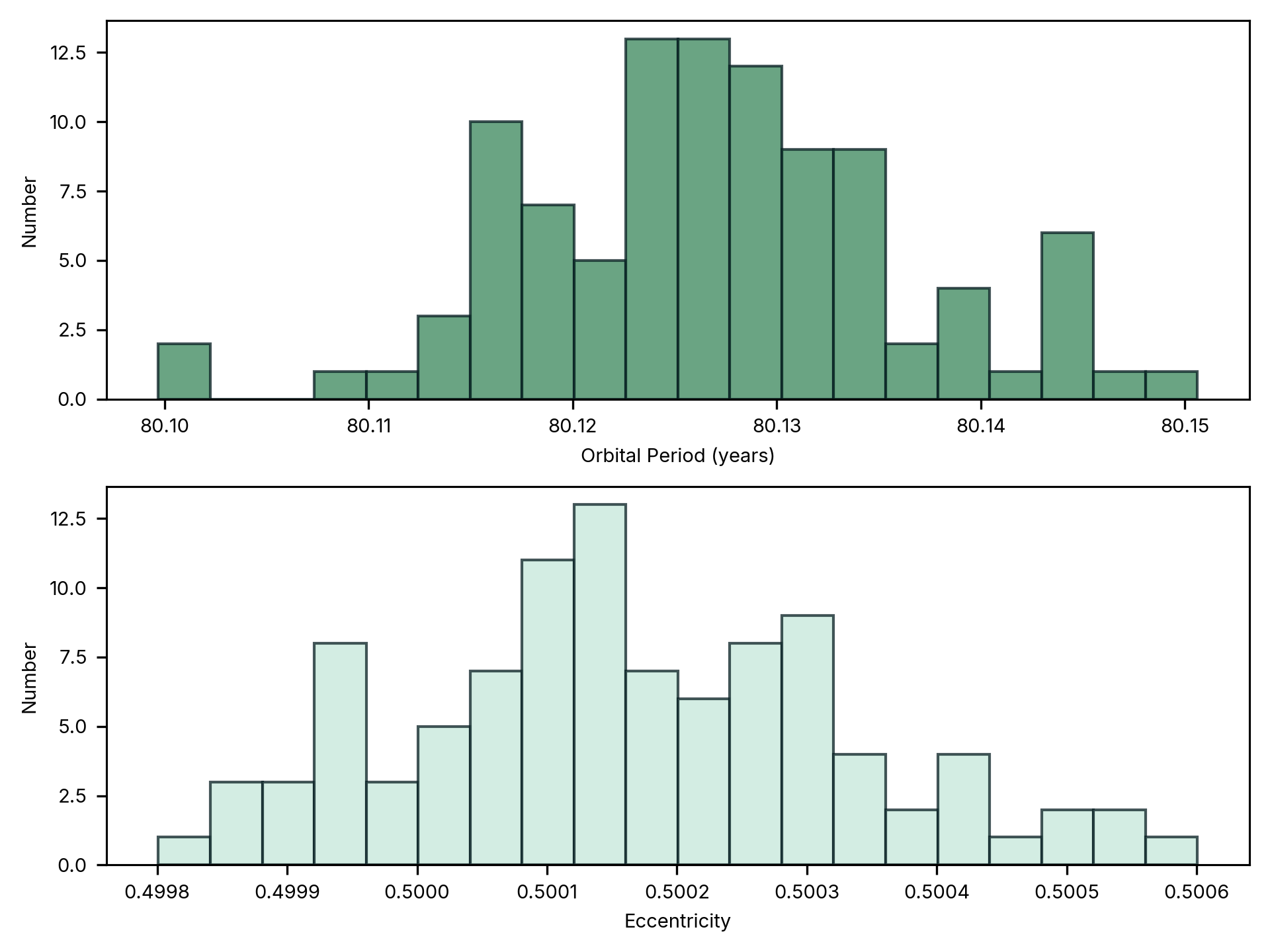}
    \caption{Histograms showing the orbit posteriors for 70 Oph.}
    \label{fig:70OphOrbitHists}
\end{figure}

\editsone{
A visual representation of 100 random draws from the range of posteriors for the orbit of the 70 Oph system derived in \citet{Li2026} is displayed in Figure \ref{fig:70OphOrbitPlot}. 
Likewise, these draws are displayed as a histogram in Figure \ref{fig:70OphOrbitHists}. 
We summarize the adopted physical properties, orbital properties, and habitable zones for the 70 Oph system in Table \ref{tab:70OphParams}.}
 \\ \\ \\ \\

\subsection{\texorpdfstring{$\gamma$ Leonis}{gamma Leonis}}
$\gamma$ Leonis (hereafter referred to as $\gamma$ Leo) is a binary composed of two red giants\footnote{$\gamma$ Leo is commonly identified in the double star literature as STF 1424AB, ADS 7724AB, WDS J10200+1950AB, and HIP 50583. $\gamma$ Leo A is HD 89484 and $\gamma$ Leo B is HD 89485.}. 

\citet{Romanenko2014} determined orbital solutions for the $\gamma$ Leo system using the Apparent Motion Parameters method, which combines astrometric measurements along short arcs, Hipparcos parallaxes \citep{HIPPARCOS1997}, and single-epoch radial velocity differences to constrain the orbital geometry. 
For our subsequent analysis, we use the E1 ($\beta~=~+38^{\circ}$) orbit posterior from Table 4 of \citet{Romanenko2014} for the binary semi-major axis, eccentricity, inclination, longitude of ascending node, and argument of periapsis. 
For the physical properties of the stars ($M_{*},R_{*},$\editsone{$L_{*}$},$T_{\rm eff}$), we adopt the values from \citet{Takeda2023}.

$\gamma$ Leo A has one confirmed planet (Ab) and one candidate planet. 
For these planets, we adopt the orbital properties and mass estimates from \citet{Han2010}. The astrometric signal due to this companion was also detected by \citet{Reffert2011}, but with a very large $\chi^2$. \editsone{Since the additional planet candidate around $\gamma$ Leo A with a proposed orbital period of $\sim1340$ days \citep{Han2010} has not yet had a firm detection, the estimates for the orbit and mass of the planet are currently imprecise.}

\editstwo{Given that the stars in the $\gamma$ Leo system are red giants, it is important to note that their habitable zones are actively evolving as the stars undergo post-main-sequence evolution \citep{Ramirez2016}. In contrast to a single-star system, this outward migration does not bring previously cold planets into the habitable zone. Instead, for this binary system, the habitable zone would have moved into a region that was likely dynamically unstable to planet formation during the system's main-sequence lifetime, given the binary's high eccentricity. Consequently, it is unlikely that planets formed in the current habitable zone in the first place. Furthermore, even were it possible for stable planets to survive in the habitable zone, they likely would not have remained there long enough to develop life under standard assumptions.}

\begin{deluxetable}{lcc}[htb!]
\tablecaption{Properties of the $\gamma$ Leo System \label{tab:GamLeoParams}}
\tablehead{\colhead{Parameter} & \colhead{Value} & \colhead{Source} }
\startdata
\hline
\multicolumn{3}{c}{$\gamma$ Leo A Properties}\\
\hline
$\mathrm{M}_*$ (\Msun)  & $1.66\pm0.14$ & \citet{Takeda2023} \\
R$_*$ (\Rsun) & $26.08\pm0.79$ & \citet{Takeda2023}\\ 
L$_*$ (\Lsun) & $251$ & \citet{Takeda2023}\\
$T_{\rm eff}$ (K) & $4457\pm23$ & \citet{Takeda2023}\\
Inner HZ Radius (AU) & $12$ & This Paper\\
Outer HZ Radius (AU) & $29$ & This Paper\\
\hline
\multicolumn{3}{c}{$\gamma$ Leo B Properties}\\
\hline
$\mathrm{M}_*$ (\Msun) & $1.55\pm0.08$ & \citet{Takeda2023} \\
R$_*$ (\Rsun) & $10.55\pm0.29$ & \citet{Takeda2023}\\
L$_*$ (\Lsun) & $63$ & \citet{Takeda2023}\\
$T_{\rm eff}$ (K) & $4969\pm15$ & \citet{Takeda2023}\\
Inner HZ Radius (AU) & $6$ & This Paper\\
Outer HZ Radius (AU) & $14$ & This Paper\\
\hline
\multicolumn{3}{c}{$\gamma$ Leo Ab Properties}\\
\hline
$M_p\sin(i_p)$ $(M_{Jup})$  & $10.7$ & \citet{Takeda2023} \\
e & $0.144\pm0.046$ & \citet{Han2010}\\
a (AU) & $1.19\pm0.02$ & \citet{Han2010}\\
$\omega$ (deg) & $206.7\pm9.2$ & \citet{Han2010}\\
\hline
\multicolumn{3}{c}{$\gamma$ Leo A Candidate Planet Properties}\\
\hline
$M_p\sin(i_p)$ $(M_{Jup})$  & $2.14$ & \citet{Han2010} \\
e & $0.13$ & \citet{Han2010}\\
a (AU) & $2.6$ & \citet{Han2010}\\
\hline
\multicolumn{3}{c}{$\gamma$ Leo Stars' Orbital Properties}\\
\hline
$P_{orb}$ (yr)  & $554\pm27$ & \citet{Romanenko2014} \\
a (AU) & $67\pm9$ & \citet{Romanenko2014} \\
e & $0.90\pm0.03$ & \citet{Romanenko2014}\\
i (deg) & $49\pm6$ & \citet{Romanenko2014}\\
$\omega$ (deg) & $309\pm6$ & \citet{Romanenko2014}\\
$\Omega$ (deg) & $347\pm9$ & \citet{Romanenko2014}\\
\enddata
\end{deluxetable}

\begin{figure}
\centering
\includegraphics[width=0.92\linewidth]{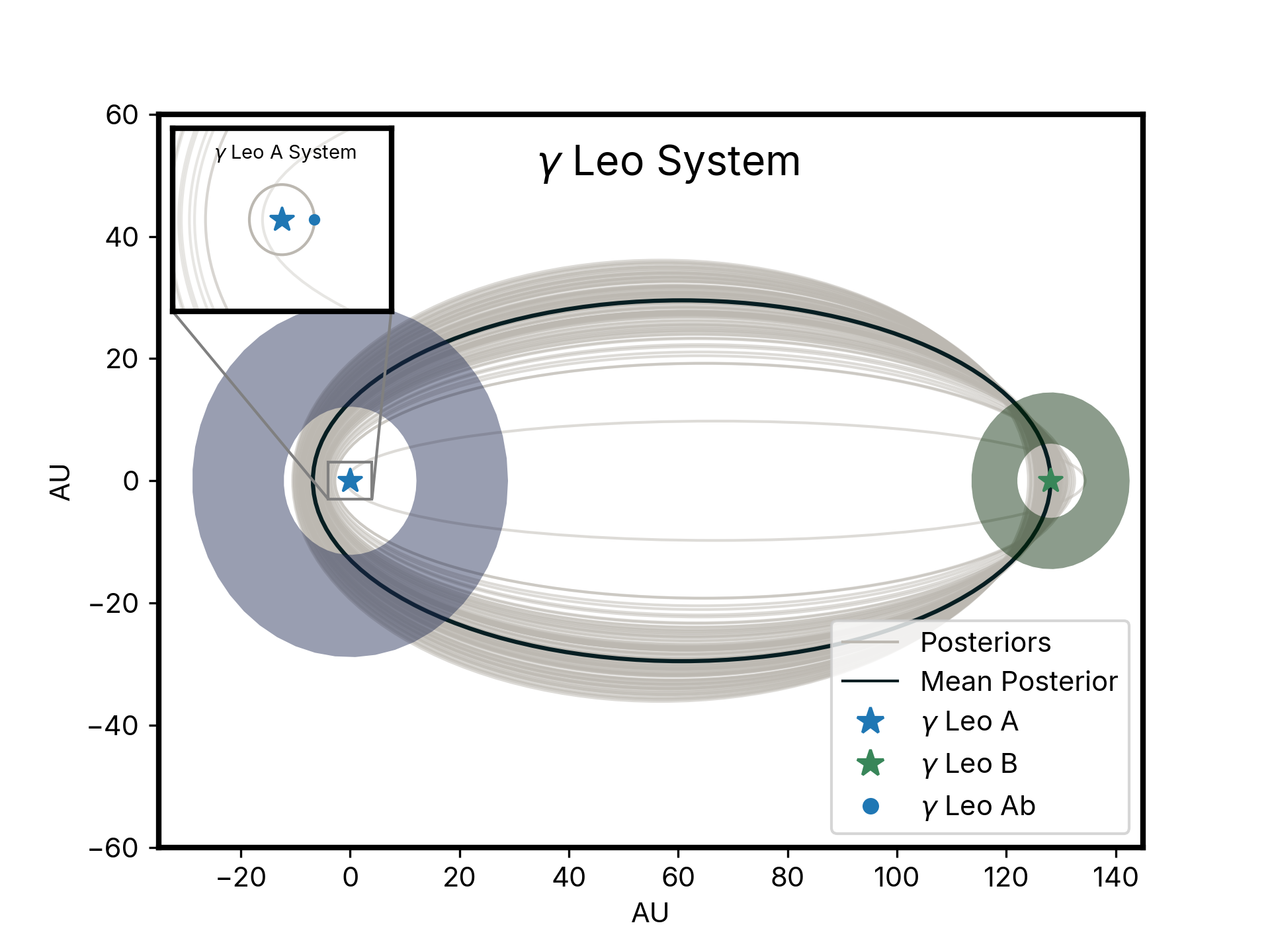}
\caption{A plot showing a representative sample of orbit posteriors for the $\gamma$ Leo system. We show $\gamma$ Leo A fixed at the origin, with the habitable zone of $\gamma$ Leo A shaded blue and the habitable zone of $\gamma$ Leo B shaded green. An inset of the $\gamma$ Leo A system shows the close orbit of $\gamma$ Leo Ab in more detail. We do not picture \editsone{the additional $\sim1340$ day candidate planet around} $\gamma$ Leonis A due to the unlikely nature of its existence based on our simulations. \editsone{As for previous orbit plots, the apparent overlap with the habitable zone of $\gamma$ Leo B is largely a consequence of the visualization.}}
\label{fig:GamLeoOrbitPlot}
\end{figure}

\begin{figure}
    \centering
    \includegraphics[width=0.92\linewidth]{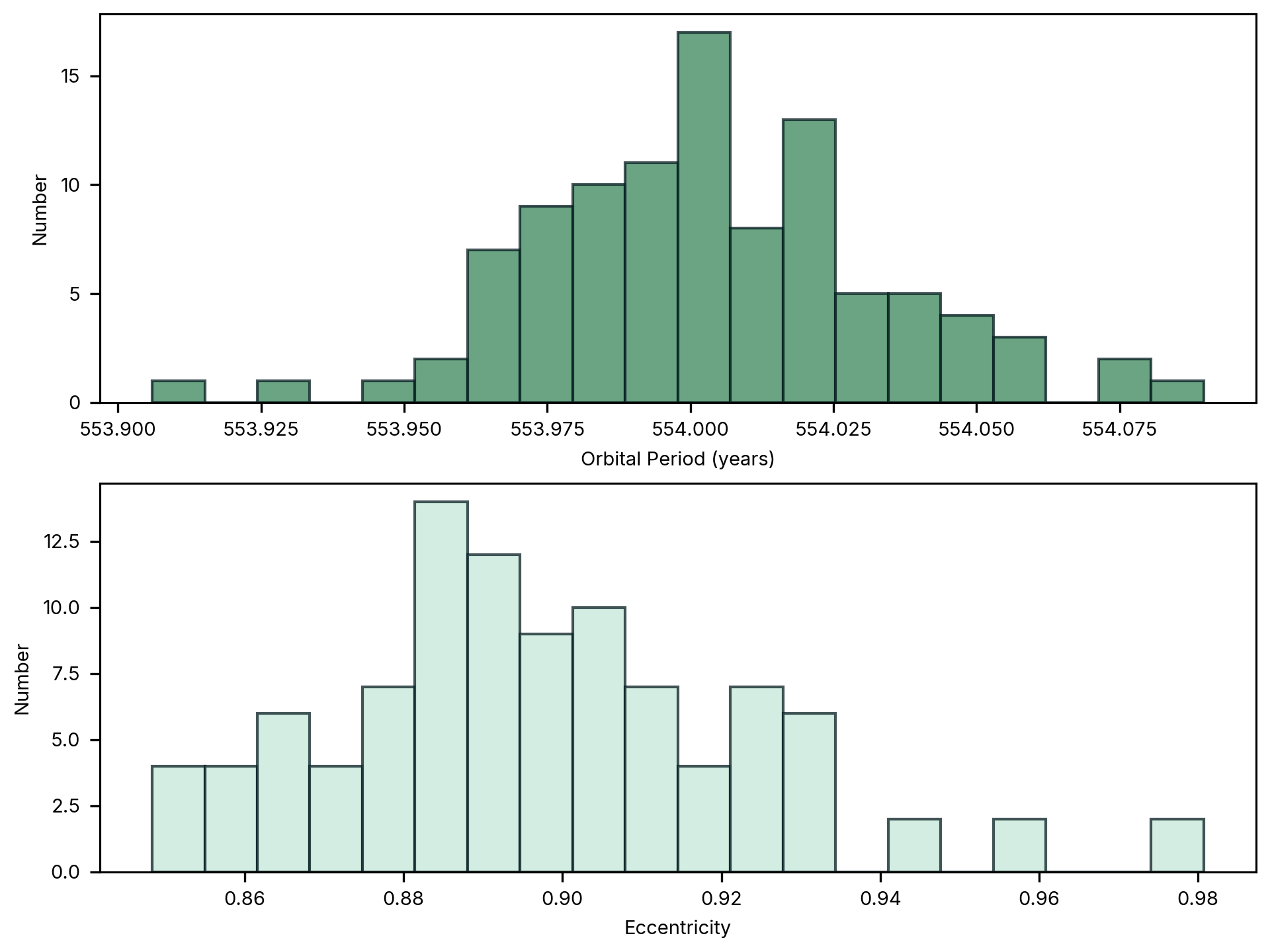}
    \caption{Histograms showing the orbit posteriors for the $\gamma$ Leo System.}
    \label{fig:GamLeoOrbitHist}
\end{figure}

We summarize the physical properties, orbital properties, and habitable zones for the $\gamma$ Leo system in Table \ref{tab:GamLeoParams}. 
\editsone
{
Figure \ref{fig:GamLeoOrbitPlot} presents a visual representation of our random samples drawn from the range of posteriors, while Figure \ref{fig:GamLeoOrbitHist} presents a histogram of these same data.
}

\section{Simulation Setup and Results} \label{sec:sims}
To assess the long-term dynamical stability and habitability of planets in each system, we performed a suite of N-body simulations using the known system parameters described in the previous section.
We use these simulations to determine whether test planets placed in the habitable zones of the component stars could remain both dynamically stable and within habitable insolation limits over extended timescales.

\subsection{Simulation Architecture}
For our simulations, we used the \editsone{IAS15} integrator from REBOUND \citep{rebound, reboundias15}. For each simulated system, we began by adding the primary star, 
\editsone{sampling randomly from the reported solution for the binary semi-major axis (a) and eccentricity (e)}. Once our stars were modeled, we added any known or theorized planets using their published parameters.

With the known or theorized bodies in each system modeled, we then injected test particle planets in the habitable zones \editsone{(defined by Equation \ref{eq:Bolmont2016})} of the host stars. 
One hundred test planets were injected in the habitable zone of each star, linearly spaced in semi-major axis \editsone{$a_p$} with the first planet injected at the innermost radius of the habitable zone and the final planet injected at the outermost radius of the habitable zone. The planets were modeled as non-interacting, massless test particles.  
Each planet's inclination \editsone{is initialized around a specified inclination angle relative to the orbit of the binary (either 0 degrees or 45 degrees), adding a random offset drawn between 0 and 0.6 degrees.} 
\editsone{For the purposes of this paper, we refer to a planet's inclination as inclination with respect to the binary orbital plane, rather than solely the inclination of the planet with respect to the equator of its host star.}
Each planet's eccentricity was generated from a uniform distribution with a lower bound of 0 and an upper bound of 0.01. Each planet's argument of periapsis, longitude of ascending node, and true anomaly were generated from a uniform distribution with a lower bound of 0 and an upper bound of $2\pi$.

Once the planets were injected, we allowed the simulations to integrate for \editsone{1 Myr, with time-steps set automatically by the IAS15 integrator}. 
\editsone{To confirm that our integration length is sufficient to capture the effect of Kozai-Lidov cycles in the 45 degree mutual inclination cases, we computed the Kozai-Lidov timescale for each system for the outermost orbit in the habitable zone of each star using Equation 42 in \citet{Antognini2015} \citep[see also][]{Naoz2016},}
\begin{equation}
    t_{KL}\simeq \frac{8}{15\pi} \left( 1+\frac{m_1}{m_3} \right) \left( \frac{P^2_{out}}{P_{in}}\right) \left( 1-e^2_{out}\right)^{3/2}
\end{equation}
\editsone{where $m_1$ is the mass of the primary, $m_3$ is the mass of the stellar companion, $P_{out}$ is the orbital period of the binary, $P_{in}$ is the period of the outermost planet in the habitable zone, and $e_{out}$ is the eccentricity of the binary.  }
\editsone{Using the parameters given in Section \ref{sec:data}, we find Kozai-Lidov timescales of $\sim2-5 \times 10^4$ years for 36 Oph, $\sim10^3 - 10^4$ years for 70 Oph, and $\sim1 \times 10^4$ years for $\gamma$ Leo. Our simulation length of 1 Myr will capture many ($>10$) Kozai-Lidov cycles for all draws from our orbit posteriors for all three systems.
}

In the following subsections, we describe the \editsone{unique aspects of the} simulations used to assess the unique \editsone{habitability} of each stellar multiple. 

\subsubsection{36 Oph}
\editsone{We ran four sets of 100 simulations for 36 Oph: }\editstwo{[1] one with the injected habitable planets at an inclination of 0 degrees (coplanar with the orbit of the binary) around 36 Oph A, [2] one with the injected planets at an inclination of 0 degrees around 36 Oph B,  [3] one with the injected planets at an inclination of 45 degrees (relative to the orbit of the binary) around 36 Oph A, and [4] one with the injected planets at an inclination of 45 degrees around 36 Oph B. 
Hereafter, we refer to the simulations where planets were injected at an inclination of 0 degrees as the coplanar case, and simulations where planets were injected at 45 degrees as the non-coplanar case (both for this and other systems).}

\editsone{For each set of 100 simulations, we drew orbits from the solution derived by \citet{NEWORBITS}. We treated each parameter independently and drew values from normal distributions centered on the reported best-fit values, with a standard deviation corresponding to the reported uncertainty. }

\subsubsection{70 Oph}
\editsone{Similar to 36 Oph, we ran four sets of 100 simulations for 70 Oph. 
We ran two coplanar cases (one for 70 Oph A and one for 70 Oph B), and two non-coplanar cases (one for 70 Oph A and one for 70 Oph B).}

\editsone{For each set of 100 simulations, we drew orbits from the solution found in \citet{Li2026}. We treated each parameter independently and drew values from normal distributions centered on the reported best-fit values, with a standard deviation corresponding to the reported uncertainty.}



\subsubsection{\texorpdfstring{$\gamma$ Leo}{gamma Leo}}

\editsone{
For the $\gamma$ Leo system, we only performed two sets of simulations, both non-coplanar cases for $\gamma$ Leo A and B. 
We chose not to simulate the coplanar cases for either system due to the architecture of the system, in which the orbits of the stars physically cross the habitable zones, and would physically interact with the test particles.
Thus, we consider the non-coplanar case to be the best case scenario for survival of the test particles, which is behavior that was displayed in our preliminary simulations.}

\editsone{For each set of 100 simulations, we drew orbits from the solution found in \citet{Romanenko2014}. We treated each parameter independently and drew values from normal distributions centered on the reported best-fit values, with a standard deviation corresponding to the reported uncertainty.}

\editsone{Regarding planet $\gamma$ Leo Ab and the additional planet candidate around $\gamma$ Leo A, we include $\gamma$ Leo Ab in our simulations, but not the additional planet candidate.
For $\gamma$ Leo Ab, we assign it the best-fit orbital parameters derived in \citet{Han2010}, using their semi-major axis, eccentricity, and argument of periapsis. We generate inclination from a uniform distribution with a width of 0.6 degrees, initialized in the same plane containing the test particle planets}, and longitude of ascending node and mean anomaly from a uniform distribution of 0 to $2\pi$.

\editsone{In this work, we neglect the $\sim1340$ day planet candidate around $\gamma$ Leo A due to its existence not yet being confirmed and due to the candidate being ejected in every preliminary simulation we ran.}



\subsection{Assessing the Habitability Of Simulated Planets}
For binary stars separated by several hundred AU, the secondary star's insolation has a negligible impact on a planet’s climate \citep{Kaltenegger2013}. At these distances, the primary threats to planetary habitability are dynamical: (1) perturbations from the companion star can induce large oscillations in a planet’s orbital elements, potentially driving it out of the habitable zone, and (2) long-term gravitational interactions may destabilize the planet’s orbit entirely, leading to ejection from the system.

We classified the planets in each of our simulations as either (a) \editsone{\textit{dynamically unstable}} if they collided with either star or attained an unbound orbit, (b) \textit{uninhabitable} if the surface flux precluded habitable conditions at any time, or (c) \textit{habitable} if the planet continually had a surface flux allowing liquid water.

To differentiate between category (b) \textit{uninhabitable} and (c) \textit{habitable}, we found the insolation flux on each planet and determined if the computed value corresponded to habitable conditions. 
\editsone{At each timestep, the flux on each particle was computed following the methods outlined in Section \ref{sec:habit}, using the $a_p$ and $e_p$ values at that simulation step.}
For a planet to be categorized as habitable, we required it to remain within the habitable boundaries for the entire lifetime of the simulation, a state called the Permanently Habitable Zone \citep[PHZ;][]{Eggl2013}. 
Any planet that \editsone{did not meet this criteria} was designated uninhabitable.

\editsone{For a planet to be categorized \editstwo{in the \textit{dynamically unstable} category}, it must have either been ejected from the system or have passed within the Roche limit of its host star. We consider a planet to have been ejected if it has attained a semi-major axis greater than two times the outer edge of its host star's habitable zone or if it has attained an eccentricity of greater than 1. The Roche limit $d$ is computed by}:
\begin{equation}
    \begin{aligned}
    d = 2.44R_\star \left(\frac{\rho_\star}{\rho_{P}} \right)^{\frac{1}{3}},
    \end{aligned}
\label{eq:RocheLimit}
\end{equation}
\editsone{where \editstwo{$R_\star$ is the radius of the host star, and $\rho_\star$ and $\rho_P$} are the densities of the stars and planets, respectively. We assume the density of the planet is 5.5 $\text{g}/\text{cm}^3$ (the bulk density of the Earth).}



\subsubsection{36 Oph}
\begin{figure}
\centering

\includegraphics[width=\linewidth]{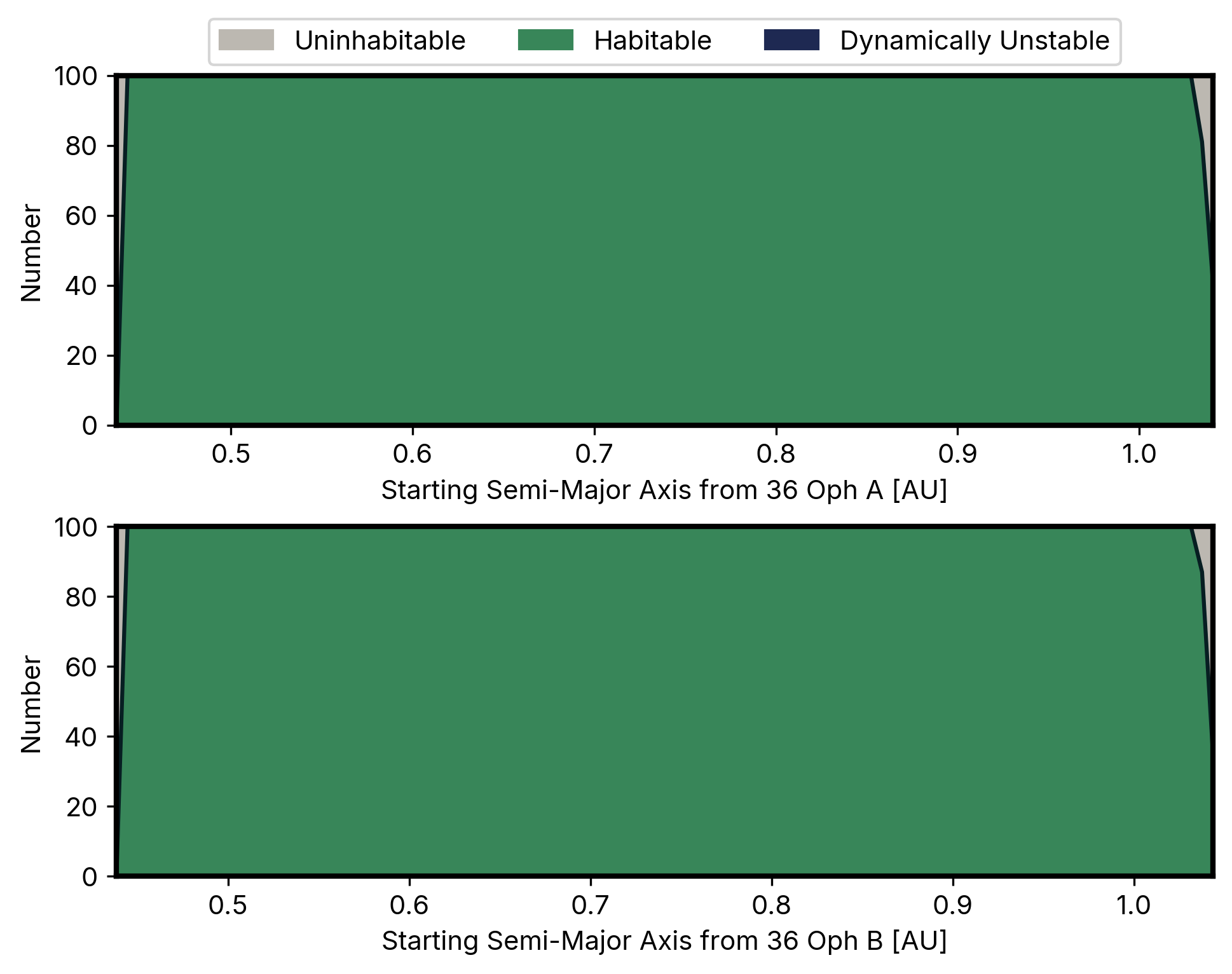}
\panelcaption{(a) Coplanar Case}

\includegraphics[width=\linewidth]{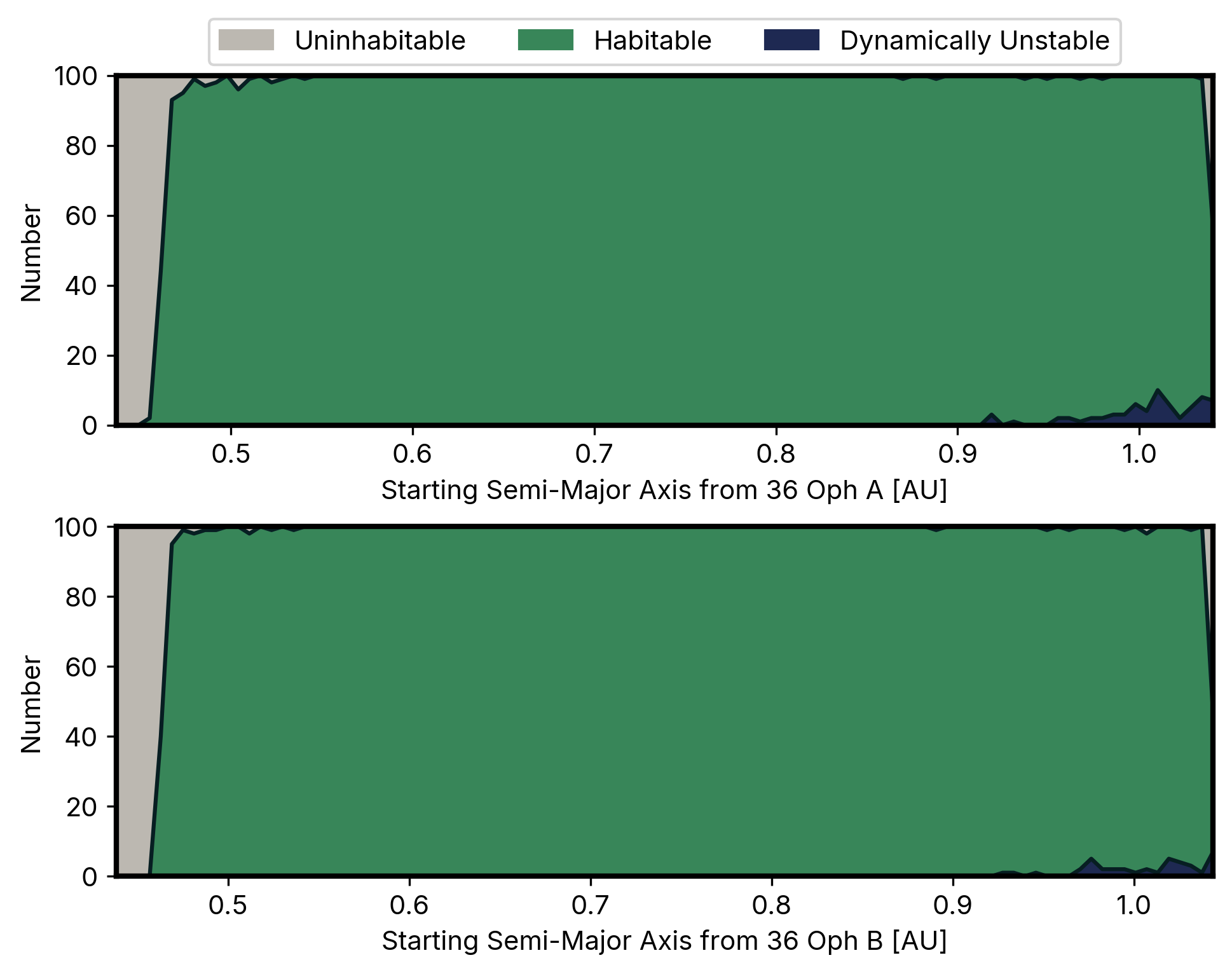}
\panelcaption{(b) Non-Coplanar Case (45 Degree mutual inclination between the binary orbit and the planet orbits)}

\caption{Plots showing the \editsone{outcomes from our simulations for 36 Oph, which include dynamical instability, habitability for the simulation duration, and uninhabitability. Panel (a) shows the coplanar case, while panel (b) shows the 45 degree mutual inclination case. Each panel contains two subpanels. The top subpanel shows survival and habitability for 36 Oph A, while the bottom subpanel shows the same for 36 Oph B.}}
\label{fig:36OphSurvivalPlots}

\end{figure}

\editsone{The results of our 36 Oph simulations are shown in Figure \ref{fig:36OphSurvivalPlots}.}
Across all our simulations, 36 Oph retained the majority of its planets within its habitable zones. Planets near the inner \editsone{or outer edges} of the habitable zone were the only planets that consistently became uninhabitable, because \editsone{the excitation of their orbital eccentricities by interactions with the binary companion caused their surface fluxes to vary beyond the limits conducive to habitability. The injected planets around 36 Oph A experienced a greater rate of dynamical instability than planets around 36 Oph B; however, ejection rates around both stars were low.}
\editstwo{At its maximum, ejection rate was only 10\% of planets around 36 Oph A and 5\% of planets around 36 Oph B, and this was only the case at the far edge of the habitable zone in the non-coplanar case. Likewise, the rate of planets going uninhabitable was only high near the close edge of the habitable zone to the stellar host, with 100\% of planets with semi-major axes less than 0.449 AU going uninhabitable around 36 Oph A and those with semi-major axes less than 0.456 AU going unstable around 36 Oph B. Once a planet is outside these ranges farther from the star, the rate of uninhabitability drops rapidly to between 0-2\%.}


Our simulation suite demonstrates that both stars in the 36 Oph A/B binary can easily host stable planets in their habitable zones.

\subsubsection{70 Oph}
\begin{figure}
\centering

\includegraphics[width=\linewidth]{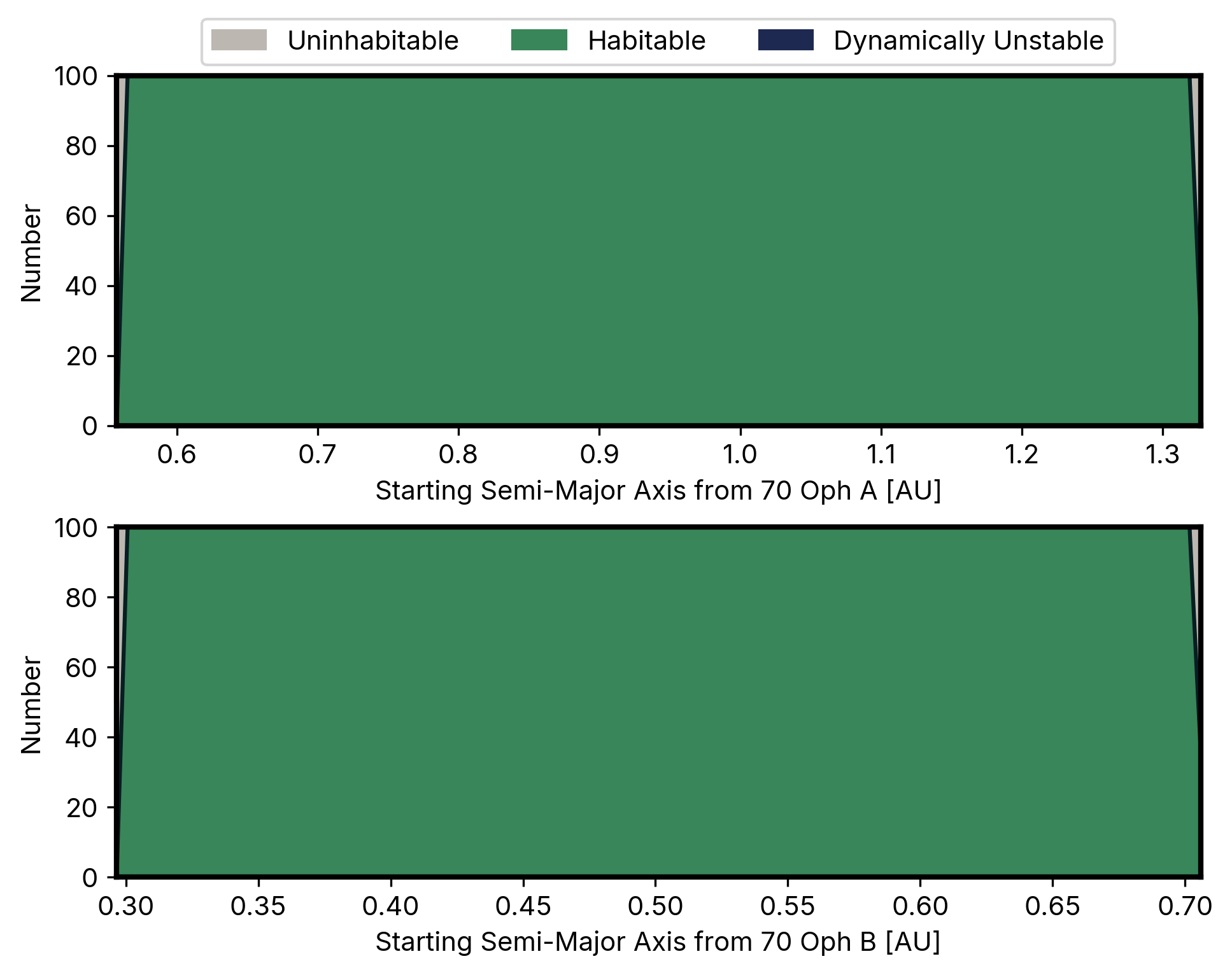}
\panelcaption{(a) Coplanar Case}

\includegraphics[width=\linewidth]{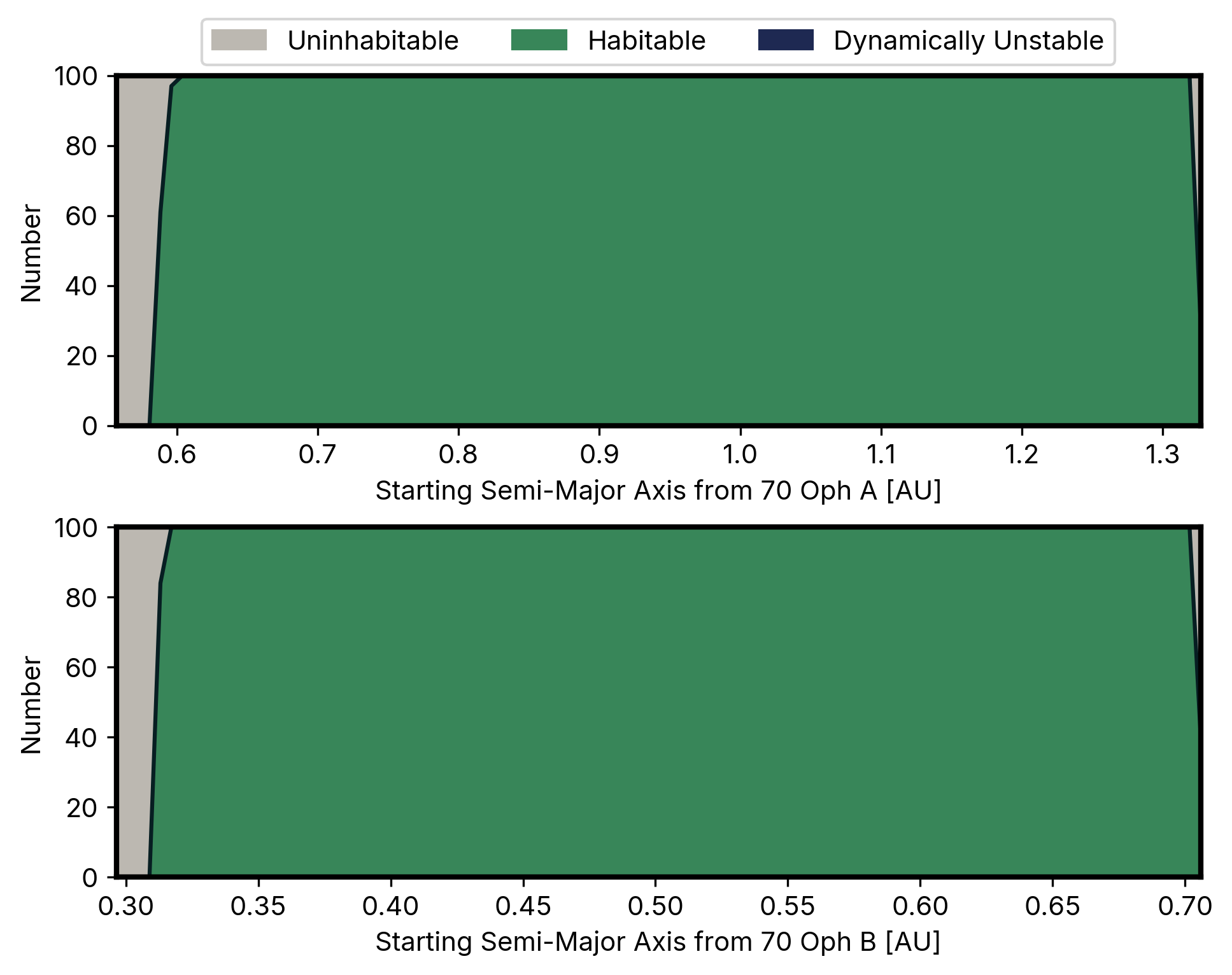}
\panelcaption{(b) 45 Degree Non-Coplanar Case}

\caption{Plots showing the ejection, uninhabitability, and habitability for the 70 Oph system per our simulations. In each panel, the top \editstwo{sub}panel shows survival and habitability for 70 Oph A, while the bottom \editstwo{sub}panel shows it for 70 Oph B.}
\label{fig:70OphSurvivalPlots}
\end{figure}

\editsone{The results of our 70 Oph simulations are shown in Figure \ref{fig:70OphSurvivalPlots}.}
\editsone{Both the coplanar case (Figure \ref{fig:70OphSurvivalPlots} a) and the 45 degree misaligned case (Figure \ref{fig:70OphSurvivalPlots} b) retained the majority of the planets in the habitable zone. However, in the latter case, Kozai oscillations did cause significant enough perturbations in the eccentricities of the test particle planets to sometimes cause them to become uninhabitable, particularly at the inner boundary of the habitable zone\editstwo{, roughly between 0.565-0.580 AU for 70 Oph A and 0.300-0.308 AU for 70 Oph B. }.} 
The most favorable orbits for habitable planets in the 70 Oph system are roughly coplanar to the plane of the binary. In that plane, habitable planets survive and maintain appropriate long-term averaged flux levels to support liquid surface water. 

\editsone{
We can also compare our results for 70 Oph to Equation 1 of \citet{Holman1999}, which estimates a critical semi-major axis beyond which planets in binary stellar systems are expected to be ejected by their hosts due to dynamical interactions. This equation can be applied to binaries with $e<0.8$ and is reproduced below:}
\begin{equation}
    \begin{aligned}
    a_c & = [(c_1)+(c_2)\mu\\
    & \ \ \ \ +(c_3)e+(c_4)\mu e\\
    & \ \ \ \ +(c_5)e^2+(c_6)\mu e^2)]a_b\ .
    \end{aligned}
\label{eq:Holman1}
\end{equation}
\editstwo{We use this equation in conjunction with the coefficients derived in Table 4 of \citet{Quarles2020}, utilizing the sets of coefficients provided for the 0 degree and 45 degree cases.}
\editstwo{\editsone{Applying this expression, we obtain a critical semi-major axis for stable planetary orbits of $3.007\pm1.319$ AU for the coplanar case, and $1.370\pm0.462$ AU for the 45 degree case.}}
All our planets orbit well within that predicted AU limit, as the farthest outer edge of the habitable zone for either star in the system is \editsone{1.32} AU. 
Thus, our results \editsone{on the dynamical stability align well with the expectations of \citet{Holman1999}, as test particles in the habitable zone are expected to be almost globally dynamically stable.}\editstwo{ Indeed, as shown in Figure \ref{fig:70OphSurvivalPlots}, our simulations indicate that the primary cause of non-habitable conditions is periodic variations in the incident stellar flux that drive the planet outside the habitable zone, rather than dynamical instability of the planet itself. Such variations are more common if the planets have a higher amount of mutual inclination relative to the binary orbit. }
\editsone{As a result, similarly to the $\alpha$ Centauri A system, the habitable zone of 70 Oph can easily support dynamically stable planets. Further astrometric and radial velocity (RV) observations of these systems might discover stable planets, as was recently done for $\alpha$ Cen A \citep{Beichman2025, Sanghi2025}.}

\subsubsection{\texorpdfstring{$\gamma$ Leo}{gamma Leo}}
\editsone{We ran \editstwo{two suites} of simulations for $\gamma$ Leo: \editstwo{one} where confirmed planet $\gamma$ Leo Ab is present and the test particles orbit their host at an inclination of 45 degrees, \editstwo{and another where confirmed planet $\gamma$ Leo Ab is present, the candidate at $\sim1340$ days is present, and the test particles orbit their host at an inclination of 45 degrees. }}

The habitable zone of $\gamma$ Leo, unlike the previous two systems tested, is incredibly dynamically unstable, \editstwo{with all of the test particle planets around both stars being ejected.} 
Under no circumstances does a\editstwo{n injected potentially habitable} planet in the $\gamma$ Leo system remain habitable, regardless of its host. Additionally, we find that, \editstwo{ while $\gamma$ Leo Ab survives in roughly 90\% of our simulations,} the potential \editsone{additional planet candidate} struggles to survive. \editstwo{Across our simulations where the unconfirmed planet candidate was present, it never remained stable. Additionally, it did not have a significant impact on the survival of $\gamma$ Leo Ab, nor did it have a significant impact on the survival of the test particles.}

In addition to the dynamical challenges for habitable planets, the stars in the $\gamma$ Leo system are red giants that have evolved off the main sequence \citep{Takeda2023}, meaning their habitable zones have evolved as well \citep{Ramirez2016}. This makes it unlikely that any planet in this system resides in the PHZ. 

We chose not to test the coplanar case for the $\gamma$ Leo system, as the orbit crossing between the stellar binary and the test particle planets would only decrease the stability in this system. 

\section{Discussion} 
In this work, we used numerical simulations to assess the long-term habitability prospects for planets in three nearby stellar multiple systems: 36 Oph, 70 Oph, and $\gamma$ Leo. 
\editsone{Our simulations show that 36 Oph and 70 Oph can support long-term, stable habitable zones around both stars, making them strong candidate\editstwo{s} for future planet searches. Planets are more often stable when in coplanar orbits with the plane of the binary; planets with inclined orbits relative to the binary plane are more likely to have their habitable conditions destabilized by Kozai–Lidov cycles. In contrast, $\gamma$ Leo shows severe dynamical instability, preventing habitable planets from persisting under current orbital solutions, and is likely not a good target for searches for habitable planets. 
}

\editsone{Recently, the topic of planets in the habitable zone around nearby stellar binaries has received renewed attention due to the evidence \citep{Beichman2025, Sanghi2025} supporting the existence of a planet in the habitable zone of $\alpha$ Cen A.}

\label{sec:discussion}
\subsection{Habitability and Observing Prospects}
In this section, we interpret the simulation results in terms of the dynamical stability and habitability prospects of the systems studied and outline the prospects for future observational constraints in each case.

\subsubsection{36 Oph}
Our simulation results indicate that habitable planets around both stars in the 36 Oph system could very likely have survived the early development of the system and would continue to be habitable to the present day. As such, we recommend that this system be considered for future searches for habitable planets. Though planets around 36 Oph B \editsone{become dynamically unstable slightly} more frequently than those around 36 Oph A, both systems have a robust region where planets remain consistently dynamically stable and habitable (as assessed by their computed surface flux). Around both stars, planets that are \editsone{at the edges of their host star's habitable zone}  are more likely to become uninhabitable, likely due to eccentricity oscillations that push the computed flux beyond the  limits of the habitable zone. We could expect to find planets in most parts of the habitable zones of the system.

\editsone{
Though the orbits for this system have varied in the literature, we believe future observations constraining the orbit will refine the best fit orbit we use here, rather than lead to a dramatically different posterior. 
We therefore do not expect our planetary stability and long-term habitability results to change dramatically.
}

\subsubsection{70 Oph}
For the 70 Oph system, our simulations show that habitability is \editsone{somewhat dependent on the initial conditions of the planets, particularly the orbital inclination with respect to the plane of the binary.}
If a planet in the habitable zone is aligned with the orbital plane of the stars, there is a high likelihood that it remains habitable.
Were planets to be discovered via astrometry in the habitable zone of 70 Oph A by future missions such as SHERA or HWO, this dynamical argument (that high misalignments relative to the stellar orbit will not permit dynamically stable planets to survive in the habitable zone) could allow a constraint on such a planet's inclination and subsequently its mass. 

The entirety of the orbit of 70 Oph has been observed, so it is well-defined. We therefore do not expect our planet stability and long-term habitability results to change dramatically even as additional observations further constrain the orbit of the stellar binary.

\subsubsection{\texorpdfstring{$\gamma$ Leo}{gamma Leo}}
The $\gamma$ Leo system, on the other hand, appears \editstwo{to host a habitable zone} far too unstable to host planets in the habitable zone of either star, even with a non-coplanar geometry that could avoid orbit crossing. 
It is the most eccentric of the three systems studied here, with an eccentricity of 0.90 (Table \ref{tab:GamLeoParams}), and as shown in the schematic orbit plot in the bottom panel of Figure \ref{fig:GamLeoOrbitPlot}, all binary orbits physically intersect with the habitable zone.
Additionally, our simulations indicate that the system is inhospitable to planets outside the habitable zone as well: \editstwo{the candidate planet at $P\sim1340$ days would not be dynamically stable, making it likely the signal is due to stellar activity or some other source rather than a planet candidate.}
\editstwo{Our results here are consistent with the expectation from the empirical stability limits \citep[e.g.,][]{Holman1999, David2003, Quarles2020}, while going beyond the formal range of those formulas and confirming that the habitable zone is unstable for the sampled posterior.}

Given the current posterior and the advanced ages of the stars, we are not able to recommend the $\gamma$ Leo system for future searches for habitable planets. For this reason, follow-up observations of $\gamma$ Leo should focus on refining the binary orbit and determining planetary properties. \editstwo{More precisely measured parameters of the existing gas giant $\gamma$ Leo A b could provide insights towards the formation processes relevant in this high-eccentricity binary system \citep[as done in, for example,][]{Stegmann2026}.}


\subsection{The Effect of Unseen System Components}
It is possible that unseen system components remain in 36 Oph, 70 Oph, and $\gamma$ Leo. For instance, 36 Oph C (though not explored in this study) may host a planetary companion, as suggested by an observed astrometric acceleration \citep{Painter2025}. While our focus is on potentially habitable planets, the presence of additional, undetected companions around any of the stars considered in this work could introduce further dynamical perturbations that impact orbital stability in ways we have not yet accounted for.
To fully assess the viability of the habitable zone, further observations are needed to ensure all potential sources of perturbations in the systems are accounted for.

\section{Conclusions} \label{sec:conclusion}
Our dynamic analysis of the 36 Oph, 70 Oph, and $\gamma$ Leo systems provides insight into the potential for habitable planets around the stars within each system.

\begin{itemize}
    \item \textbf{36 Oph}: Both stellar components can sustain stable habitable zones, making the system a strong observational target for future missions. \editsone{That said, planets in coplanar orbits are more likely to remain habitable, while inclined orbits are more likely to be disrupted by Kozai-Lidov dynamics.}
    \item \textbf{70 Oph}: \editsone{Both stellar components can sustain stable habitable zones, making the system a strong observational target for future missions. However, inclined orbits can be disrupted by Kozai-Lidov dynamics.}
    \item \textbf{$\gamma$ Leo}: It appears that the habitable zone is dynamically unstable under current orbital solutions, with neither star able to host long-term habitable planets. \editsone{Additionally, the orbit for the $\gamma$ Leo system suggests that the existence of the $\sim1340$ day planet candidate mentioned in \citet{Han2010} is incredibly unlikely due to its likely dynamical instability. The alternative hypothesis from \citet{Han2010} that this signal could instead be due to stellar activity would be more consistent with the results of our simulations.}
\end{itemize}

The procedure outlined in this paper is a framework for future analyses of other stars in binary systems to determine their potential for hosting habitable worlds.


\section*{Acknowledgments}
A.J. thanks the generous support of the Peter Livingston Scholars Program at the University of Wisconsin--Madison. 
This research has made use of the NASA Exoplanet Archive, which is operated by the California Institute of Technology, under contract with the National Aeronautics and Space Administration under the Exoplanet Exploration Program. This research has made use of NASA's Astrophysics Data System Bibliographic Services.
Part of this research was carried out at the Jet Propulsion Laboratory, California Institute of Technology, under a contract with the National Aeronautics and Space Administration (80NM0018D0004).
Part of this research was supported by NASA XRP 80NSSC25K7148.


\vspace{5mm}
\facilities{Exoplanet Archive \citep{Christiansen2025}}

\software{\texttt{REBOUND} \citep{rebound}
\texttt{matplotlib} \citep{Hunter2007},
\texttt{pandas} \citep{mckinney-proc-scipy-2010, the_pandas_development_team_2024_13819579},}

\begin{appendix}
\section{Discussion of the Literature Orbit Solutions}
\label{app:solutions}

\subsection{36 Oph}
\label{app:36oph}
\editsone{
The two orbits previously reported in the literature for 36 Oph vary significantly. 
In \citet{Irwin1996} (later cited in \citealt{Tokovinin2017}), three orbits are derived: one with just visual astrometric data, one with astrometric data and RV's, and one with the observed $\Delta V_{B-A}$.
In contrast, the derived orbits in \citet{Izmailov2025} are reported as 100 covariant orbit posteriors solely based on visual data (including recent data from Gaia), and exhibit a heavy bimodal distribution in the estimated orbital period (P) and semi-major axis (a).
One of the two modes reported by \citet{Izmailov2025} is roughly in line with the orbits derived in \citet{Irwin1996}, with $P_{orb,mean}=598.6$ yr in \citet{Izmailov2025} compared to $P_{orb}=586.9$ yr in Orbit 4 in \citet{Irwin1996}.}

\editsone{However, eccentricity (e) for these orbits differs dramatically, with $\bar{e}=0.33540$ in \citet{Izmailov2025} and $e = 0.9223699\pm0.0011$ in Orbit 4 in \citet{Irwin1996}, which have very different implications for the dynamical stability of additional bodies in the system. 
Though \citet{Irwin1996}'s use of RVs should make it a more complete solution, \citet{Izmailov2025} uses nearly 30 years of additional astrometric data.}

\editsone{In this work, we adopt the solution of \citet{NEWORBITS}, which uses both RV and astrometric data to obtain a solution with an eccentricity of $0.8999 ^{+0.0032}_{-0.0033}$. Of the previous literature solutions, this newly measured eccentricity is more in line with the \citet{Irwin1996} solution.}

\subsection{70 Oph}
\label{app:70oph}
\editstwo{The orbit of 70 Oph has been studied for more than a century through a combination of visual astrometry and radial velocity observations, starting around the 1875 periastron passage \citep[as described in][]{Batten1984}.  More modern solutions following the 1984 periastron passage \citep{Batten1984, Heintz1988, Batten1991} provided updated solutions \citep{Eggenberger2008}. Since the 1984 periastron passage, the solutions have been in good agreement on the orbital period ($P\sim 88$ years) and eccentricity ($e\sim 0.5$). More recently, \citet{Piccotti2020} computed updated stellar masses using stellar evolution tracks, yielding a primary mass of $\mathrm{M}_* = 0.896\pm0.083$ \Msun and a secondary mass of $\mathrm{M}_* = 0.782\pm0.064$ \Msun. \citet{Izmailov2025} presented a covariant astrometric orbit solution based on historical visual measurements and Gaia observations, which was in good agreement with previous solutions but did not include all available radial velocity data. }

\editstwo{In this work, we adopt the solution of \citet{Li2026}, which combines more than a century of astrometric monitoring with 27 years of archival radial velocities and new Planet Finder Spectrograph observations obtained between 2023 and 2025. This updated, joint astrometric and radial velocity analysis yields updated orbital elements and precise dynamical masses for both stellar components, reproduced in Table \ref{tab:70OphParams}. This new solution yields an improved precision on orbital parameters and stellar masses (finding a primary mass of $\mathrm{M}_* = 0.8827\pm0.004$ \Msun and a secondary mass of $\mathrm{M}_* = 0.7319\pm0.0031$ \Msun).}

\subsection{$\gamma$ Leo}
\label{app:gamleo}
\editstwo{Unlike 36 Oph and 70 Oph, there is not yet a succession of modern orbital solutions for the $\gamma$ Leo binary. The orbit we adopt in the present work was derived by \citet{Romanenko2014} using a combination of astrometry and radial velocity measurements. Their preferred E1 ($\beta=+38^\circ$; one of three solutions provided in the work, yields the binary orbital parameters used in this work and reproduced in Table \ref{tab:GamLeoParams}. The other solution for the binary orbit reported in \citet{Romanenko2014} has very similar parameters (identical orbital period with $e=0.93$ instead of our adopted $e=0.90$). }

\editstwo{The remaining system parameters are not reported with the orbital solution, and as a result we take them from additional works. We adopt the stellar masses, radii, luminosities, and effective temperatures from \citet{Takeda2023}. We note that the stellar masses obtained by the stellar evolution models of \citet{Takeda2023} give a significantly lower combined mass ($3.21 M_{\odot}$) than the earlier dynamical solutions for the mass \citep[$5.5 M_{\odot}$ from][]{Romanenko2014}. We consider the stellar evolution masses more reliable and adopt them in this work. 
For the confirmed planet $\gamma$ Leo Ab, we adopt the orbital elements from \citet{Han2010}. We also adopt the proposed orbital parameters of the additional $\sim1340$ day planet candidate from \citet{Han2010}, although we emphasize throughout this work that its existence remains unconfirmed.}

\end{appendix}
\bibliography{bibliography}{}
\bibliographystyle{aasjournal}
\end{document}